\documentclass[epj]{svjour}
\usepackage{epsfig} 
\usepackage{times}
\begin{document}
%\title{%\hfill{\tiny FZJ-IKP-TH-2005-12, HISKP-TH-05/09}\\
%}
\title{{\boldmath $K{\bar K}$} photoproduction from protons}
\author{A.~Sibirtsev\inst{1}, J.~Haidenbauer\inst{2}, 
S.~Krewald\inst{1}, U.-G.~Mei{\ss}ner\inst{1,2}
and A.W.~Thomas\inst{3}} 
\institute{Helmholtz-Institut f\"ur Strahlen- und Kernphysik (Theorie), 
Universit\"at Bonn, Nu\ss allee 14-16, D-53115 Bonn, Germany \and
Institut f\"ur Kernphysik (Theorie), Forschungszentrum J\"ulich,
D-52425 J\"ulich, Germany
\and Jefferson Lab, 12000 Jefferson Ave., Newport News, VA 23606, USA}
\date{Received: date / Revised version: date}

\abstract{We study the contribution of the Drell mechanism driven by $K^+$ and
$K^-$ exchange to the reaction $\gamma{N}{\to}K{\bar K}N$. Our calculation
implements the full $KN$ and ${\bar K}N$ reaction amplitudes 
in the form of partial wave amplitudes taken from a 
meson-exchange model ($KN$) and a partial wave analysis ($\bar KN$),
respectively. Comparing our results to data of the LAMP2 collaboration
we observe that the Drell mechanism alone cannot describe the
large $\Lambda$(1520) photoproduction rate observed experimentally. 
We argue that the discrepancy could be due to significant 
contributions from $K^\ast$ meson exchange with subsequent excitation
of the $\Lambda$(1520) resonance. 
After adding such contributions to our model a good agreement 
of the LAMP2 experiment is achieved. 
When applying the same
model to the recent SAPHIR data we find an excellent description
of the $K^+ p$ spectrum and we find evidence for a hyperon resonance 
with $M_R = 1617\pm 2\,$MeV and $\Gamma_R =
117\pm 4\,$MeV in the $K^- p$ mass distribution.}

\PACS{ {11.80.-m} {Relativistic scattering theory} \and 
{11.80-Et} {Partial-wave analysis} \and
{12.40.Nn} {Regge theory, duality, absorptive/optical models} 
\and  {13.60.Le} {Meson production} \and  
{13.60.Rj} {Baryon production}
\and  {13.75.Jz} {Kaon-baryon interactions} } 

\authorrunning{A.~Sibirtsev, J.~Haidenbauer, S.~Krewald,
U.-G.~Mei{\ss}ner and A.W.~Thomas} \titlerunning{
{\boldmath $K{\bar K}$} photoproduction from protons}

\maketitle

\section{Introduction}
The reaction $\gamma{N}{\to}K{\bar K}N$ offers an excellent opportunity
for hadronic spectroscopy. First of all, it allows access to the $KN$ system,  
a channel which has received much attention recently because the exotic 
$\Theta^+$(1540) pentaquark couples to it. With regard to that issue 
experiments were performed 
at SPring$-$8, CEBAF and ELSA using either a free or bound target 
nucleon \cite{Nakano,Stepanyan,Barth,Kubarovsky}. 
The ${\bar K}N$ system can serve as a source for
hyperon resonance spectroscopy. In fact, the most recent data
on the $\Lambda$(1520) hyperon quoted by PDG~\cite{PDG}
were obtained by the LAMP2 group \cite{Barber}
from the reaction $\gamma{p}{\to}K^+K^- p$ at photon energies
$2.8\le E_\gamma \le 4.8$~GeV. Moreover, the missing mass spectrum measured 
in $\gamma{p}{\to}K^+X$ at $E_\gamma$=11~GeV~\cite{Boyarski} 
indicates many $\Lambda$ and $\Sigma$ resonances 
and clearly illustrates the promising perspectives of the reaction 
$\gamma{N}{\to}K{\bar K}N$ 
with respect to hyperon spectroscopy. Finally, the $K{\bar K}$ 
system allows one to investigate mesonic resonances with normal and exotic 
quantum numbers. Among those are the light scalar mesons $f_0(980)$
and $a_0(980)$ whose nature is still under debate~\cite{Bugg,Amsler,Baru5}. 
 
The different baryonic and mesonic resonances that can occur in the reaction
$\gamma{N}{\to}K{\bar K}N$ induce a considerable complexibility in 
the overall data analysis. In particular, possible kinematical reflections,
but also momenta and angular cuts of the final state due to a
limited detector acceptance might generate resonance-like
structures~\cite{Byckling}, which do not correspond to genuine physical 
quantities.
Thus, it would be rather useful to construct a phenomenological model 
that allows to control and constrain the background as
much as possible. Thereby, it is crucial that one includes not only
``true'' background contributions but also available experimental 
information on already well established baryonic and mesonic resonances 
that may contribute to the final state. 
Indeed, a similar strategy was followed by Drell~\cite{Drell1},
S\"oding~\cite{Soding1} and Krass~\cite{Krass}  
for the analysis of the reaction $\gamma{p}{\to}\pi^+\pi^-p$
already a long time ago. 
In particular, S\"oding~\cite{Soding1} suggested that one should construct a
model that incorporates the Drell mechanism
using available experimental information for  
$\pi{N}{\to}\pi{N}$ elastic scattering and then apply this model
to the $\gamma{N}{\to}\pi\pi{N}$ reaction to search for new
resonances or new phenomena~\cite{Ballam,Erbe}.

The principal aim of our work is to develop a model for the 
background to the reaction $\gamma{N}{\to}K{\bar K}N$ by 
incorporating, as far as possible, all presently available 
and well established experimental results.   
To be more concrete, in the present paper
we consider the Drell mechanism for the reaction $\gamma{N}{\to}K{\bar K}N$.
This allows us to take into account the experimental 
knowledge of the reaction amplitudes in the $KN$ and ${\bar K}N$
subsystems. Specifically, we can naturally account for the 
manifestation of resonances in the $\bar KN$ and $KN$ 
channels. For the former this concerns, in particular, the 
$\Lambda$(1520) resonance. 
The results of our calculation are then compared with available 
experimental data on photoproduction of the $\Lambda$(1520) and
specifically with the $K^+p$ and $K^-p$ invariant mass spectra
reported by Barber et al. \cite{Barber}. 

Within the last year or so several papers have appeared 
that deal with the reaction $\gamma{N}{\to}K{\bar K}N$ 
and/or the photoproduction of the $\Lambda$(1520) hyperon 
\cite{Roberts,Oset,Oh,Nam1,Nam2,Grishina8,Titov8}. 
Most of those investigations were driven by the quest for the
$\Theta^+$(1540) pentaquark. Though a large variety of reaction 
mechanisms were considered none of those studies takes
into account the Drell mechanism based on the full $KN$ and
$\bar KN$ amplitudes. In fact, in some works $K$ exchange is
considered but then the $KN$ and/or $\bar KN$ amplitudes are 
approximated by tree-level resonance and/or 
t-channel meson-exchange diagrams \cite{Oh,Titov8}. \\ 
Therefore, our present paper is complementary to those other
studies and, moreover, it has the potential to provide 
additional and important information on the photoproduction
of the $K\bar K$ system. Specifically, it allows for a reliable
evaluation of the background contribution due to $K$ exchange 
that is present in all such photon induced reactions
and thus should be taken into account in the analysis of
experimental results. 

It will be shown in our analysis that the Drell mechanism alone
is not sufficient to explain the $K^+p$ and $K^-p$ invariant
mass spectra measured by the LAMP2 group \cite{Barber}. 
Specifically the latter is significantly underestimated in
the region of the $\Lambda$(1520) peak. 
This implies that the $\Lambda$(1520) photoproduction cross
section is much larger than in corresponding hadron 
induced reactions.
We then consider $K^\ast$ meson exchange in conjunction with
the excitation of the $\Lambda$(1520) resonance as
additional reaction mechanism. Within such a scenario 
it is indeed possible to achieve a satisfactory description of
the large $\Lambda$(1520) photoproduction rate as reflected in 
the $K^-p$ mass spectrum \cite{Barber}.

The paper is organized as follows.
In Section~2 we formulate our model to calculate the Drell mechanism. 
In particular we specify the $KN$ and $\bar KN$ amplitudes that we
employ in our investigation. 
Section~3 presents a comparison between available data on $K^+p$
and $K^-p$ invariant mass spectra from the reaction 
$\gamma{N}{\to}K{\bar K}N$ with our results based on the Drell
mechanism. We also consider additional contributions which we assume
to be due to $K^\ast$ meson exchange. In Section~\ref{sec:ELSA}
we apply our model to the recent SAPHIR data for $\gamma p\to
K^+ K^- p$. We demonstrate that from these data one can indeed
deduce the parameters of a hyperon resonance, probably the 
$\Lambda$(1600). A comparision with new data from the CLAS 
Collaboration is presented in Section~\ref{sec:ELSA}. 
We summarize our results with our Conclusion. 

\section{The Drell mechanism and the {\boldmath$\bar KN$} reaction}

\begin{figure}[t]
\vspace*{-8mm}
\centerline{\hspace*{7mm}\psfig{file=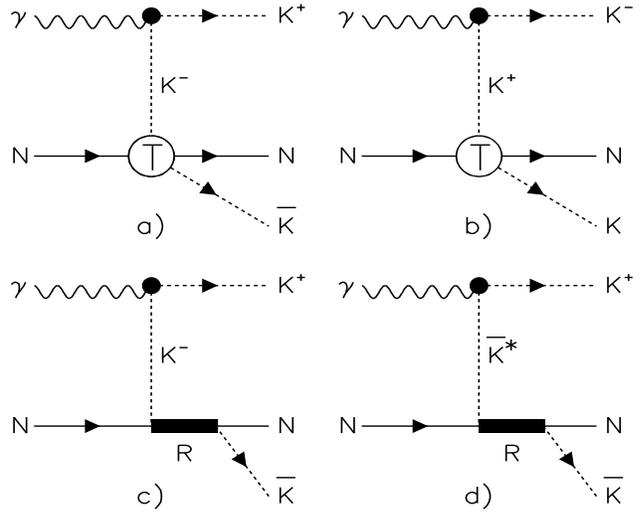,width=9.9cm,height=8.4cm}}
\vspace*{-3mm}
\caption{Diagrams for the reactions $\gamma{N}{\to}K{\bar K}N$.
The Drell mechanism for $K$ exchange with full ${\bar K}N$ and $KN$ 
resonant and nonresonant amplitudes is shown by a) and b). 
Diagrams c) and d) show $K$ and $K^\ast$ exchanges through the 
excitation of $Y$ hyperon resonances only. Note that in the Drell
mechanism the diagram c) is naturally included in the process a).}
\label{digdrell}
\end{figure}

The Drell mechanism for the reaction $\gamma{N}{\to}K{\bar K}N$ is shown 
by the two diagrams a) and b) in Fig.~\ref{digdrell}.  Only the exchanges 
of charged kaons $K^+$ and $K^-$ contribute to the reaction, since 
photons do not couple to neutral $K$ mesons. However, the rescattering
amplitude
includes both elastic scattering and charge exchange. The 
amplitude for $t$-channel $K^-$ exchange is given   
as~\cite{Drell1,Soding1,Krass}
\begin{eqnarray}
{\mathcal M}_{K^-} = -2 {\vec \epsilon_\gamma}\cdot{\vec q_1}\,
\frac{e\,T_{K^-p}(s_2,t_2)\,F(t_1)}{t_1-m_K^2},
\label{Drell_a}
\end{eqnarray}
where $\vec \epsilon_\gamma$ is the photon polarization vector, 
$\vec q_1$ is the momentum of the $K^+$ meson in the $\gamma{N}$ c.m. system,
$t_1$ is the squared four-momentum transferred between the photon and the 
$K^+$ meson and $T_{K^-p}$ is the $K^-p$ scattering or charge exchange
amplitude, which 
depends on the squared invariant mass $s_2$ of the $K^-p$ system and the
squared four-momentum $t_2$ transferred from the initial to the final proton.
Furthermore, $F$ is a form factor that accounts for the offshellness of the 
$K^-p$ scattering amplitude. This form factor is taken in the form 
proposed by Ferrari and Selleri~\cite{Ferrari2}:
\begin{eqnarray}
F(t_1)=\frac{1}{1+(m_K^2-t_1)/t_0} \ .
\label{ff}
\end{eqnarray}
We fix the cutoff parameter $t_0$ via a fit to data on the total reaction 
cross section for $\gamma{p}{\to}K^+K^-p$. The resulting value is
$t_0$=0.33 GeV$^2$, which corresponds to a cutoff mass of 
$\Lambda$ = 0.76 GeV within a standard monopole form factor. 
The $\gamma{K^+K^-}$ vertex
contains no form factor because of the Ward identity.
The amplitude for $t$-channel $K^+$ exchange has the same structure
except that then $T_{K^+p}$ enters and, of course, the form factor $F$
in the $KN \to KN$ vertex could be different as well.  
We implement current conservation by adopting the
prescription of deForest and Walecka to replace the longitudinal 
current $\hat{q} \cdot \vec{j}(\vec{q})$ by $ j^0(\vec{q})q^0/q$ \cite{dFW},
where $q$ denotes the photon momentum. 
This corresponds to the addition of an appropriate contact interaction
between the photon, kaon, and nucleon. Note, that such contact terms 
arise also in the formalism of Haberzettl \cite{haber97} which was
recently applied to the photoproduction of kaons in the vicinity of the 
pentaquark \cite{titov05}, though one must say that in the latter case 
this prescription is based on the Ward-Takahashi identity while it is
essentially ad hoc in our case. 
 \

In case of kaon exchange the elementary (${\bar K}N$ and $KN$) scattering
processes take place even farther off-shell than those in the 
reaction $\gamma{p}{\to}\pi^+\pi^-p$. Still we believe that it is a 
good working hypothesis to use the on-shell amplitudes in 
Eq.~(\ref{Drell_a}) but take into account the offshellness of the actual 
elementary reaction effectively by introducing a form factor. 
It rests on the expectation that the $s_2$ (and $t_2$) dependence of the 
elementary amplitudes, c.f. Eq.~(\ref{Drell_a}), remain largely unchanged 
when the reaction takes place off-shell. This expectation is supported by 
experience with other reactions. For example, one knows from meson-production reactions 
in $NN$ collisions ($pp \to ppx$, $x$ = $\pi$, $\eta$, $\eta'$ ...) that the energy
dependence of the reaction cross section is strongly dominated by the energy 
($s$) dependence of elastic $pp$ scattering, even though in the production
reaction the relevant $pp$ interaction in the final state takes place far 
off-shell, see, e.g., Ref.~\cite{Hanhart}. 

In the Drell formulation
all possible resonances coupled to either the $KN$ or ${\bar K}N$ channels
enter the calculations through the $KN{\to}KN$ or ${\bar K}N{\to}{\bar K}N$
scattering or charge exchange
amplitudes. Keeping in mind that the available data
on hyperon-resonance properties are quite uncertain~\cite{PDG},
calculations utilizing the experimentally available ${\bar K}N{\to}{\bar K}N$ 
amplitudes at least allow one 
to incorporate the best phenomenological information into the treatment of  
the reaction $\gamma{N}{\to}K{\bar K}N$. In that respect the Drell mechanism
provides consistency between the data on ${\bar K}N{\to}{\bar K}N$ 
elastic scattering and $K{\bar K}N$ photoproduction. Similar
considerations hold for the $KN$ system. 

Originally S\"oding~\cite{Soding1} proposed that one parameterizes the invariant 
elastic amplitude (in our case $T_{\bar K N}$ and $T_{KN}$)
directly from the data on the total cross section $\sigma_{tot}$ 
by exploiting the optical theorem, 
\begin{eqnarray}
T_{KN}(s_2,t_2)= -2iq_2\sqrt{s_2} \, \, \sigma_{tot}(s_2)\exp{(bt_2)},
\label{soding}
\end{eqnarray}
where $s_2$ is the $KN$ (${\bar K}N$) invariant mass squared, $q_2$ 
is the modulus of the c.m. momentum and $t_2$ is the squared
four-momentum transferred from the initial to the final nucleon. Here
$b$ is the exponential slope of the $t$ dependence
taken from the data. This approximation is, in principle, sufficient to calculate
the energy dependence of the reaction and the $KN$ (${\bar K}N$) 
mass distribution.
However, for an application involving cuts on certain momenta and angles one 
needs the specific dependence of the $KN$ (${\bar K}N$) amplitude on those 
quantities and then the simple $\exp{(bt)}$ ansatz is not 
adequate.

In order to circumvent this problem 
Berestetsky and Pomeranchuk~\cite{Berestetsky} and also Ferrari and 
Selleri~\cite{Ferrari} proposed that one parameterize the invariant
scattering amplitude directly from differential elastic 
scattering cross section data:
\begin{eqnarray}
|T_{KN}(s_2,t_2)|^2=64\pi \, s_2 \, q_2^2 \, \frac{d\sigma_{el}(s_2)}{dt_2}.
\label{elast}
\end{eqnarray}
At least this approximation allows one to account for many details of 
the nuclear reactions and to reconstruct general features of the 
underlying dynamics~\cite{Sibirtsev1,Sibirtsev2,Sibirtsev3}.
However, even this form is not suitable for the present case where 
two different amplitudes connected with $K^+$ and $K^-$ meson exchanges,
respectively, enter the calculation coherently. 

Thus, the
most consistent way is to use the partial wave decomposition
of the scattering amplitude given by
\begin{eqnarray}
G(s_2,\theta )=\frac{1}{q_2}\sum_l [(l+1)T^+_l(s_2)+lT^-_l(s_2)]
P_l(\cos\theta ), 
\nonumber \\
H(s_2,\theta )=\frac{\sin \theta}{q_2} \sum_l[T^+_l(s_2)-T^-_l(s_2)]
\frac{dP_l(\cos\theta)}{d\cos\theta},
\label{pwa}
\end{eqnarray}
where $G$ and $H$ are the spin-nonflip and spin-flip amplitudes \cite{Hoehler}, 
respectively,
$P_l(\cos\theta)$ are the Legendre polynomials and $\theta$ is the 
scattering angle
in the c.m. system. The partial wave amplitudes (for the $KN$ as well as
the $\bar KN$ systems) are defined as
\begin{eqnarray}
T_l^\pm = \frac{\eta_l^\pm \exp(2i\delta_l^\pm )-1}{2i},
\end{eqnarray} 
where $\eta_l^\pm$ and $\delta_l^\pm$ denote the inelasticity and the
phase shift, respectively, for the total angular momentum $J{=}l{\pm}1/2$.
The relation between the invariant $G$ and $H$ scattering amplitudes
and various scattering observables can be found, for example,  
in Ref.~\cite{Hoehler}. An application of the partial wave 
amplitudes to the analysis of the reaction $\gamma{p}{\to}\pi^+\pi^-p$ 
for the Drell mechanism contribution using
the S\"oding model is described in detail in Ref.~\cite{Ballam}.
Note that the partial wave decomposition is given in the 
isospin basis and therefore any of the final channels 
of the reaction $\gamma{N}{\to}K{\bar K}N$ can be calculated.

In our investigation we use the $KN$ amplitudes from the J\"ulich 
meson-exchange model. A detailed description of the model is given in
Refs.~\cite{Buettgen,Hoffmann}. The model yields a satisfactory 
reproduction of the available 
experimental information on elastic and charge exchange $KN$ scattering 
including angular spectra and polarization data up to an invariant mass
of the $KN$ system of $\sqrt{s_2}{\simeq}1.8$~GeV. For higher energies we 
adopt the phenomenological ansatz given by Eq.~(\ref{soding}) 
utilizing experimental data compiled in Ref.~\cite{Flaminio}. 
Let us mention in this context that the $KN$ amplitude
from the J\"ulich model was used by us recently for determining 
new limits for the $\Theta^+$ pentaquark width from the data available
for the $K^+d{\to}K^0pp$ reaction~\cite{Sibirtsev4} and also for an 
analysis of the DIANA experiment~\cite{Dolgolenko}, where 
the $\Theta^+$ pentaquark was reportedly observed
in $K^+$ meson collisions with $Xe$ nuclei~\cite{Sibirtsev5}.
Therefore, as an advantage, our approach offers the possibility for the
self-consistent inclusion of the $\Theta^+$ pentaquark via the $KN$ 
scattering amplitude, where the latter has already been developed and compared 
to other available data~\cite{Sibirtsev4,Sibirtsev5}.

The ${\bar K}N$ amplitudes are reconstructed from the result of a 
multichannel partial wave analysis (PWA)~\cite{Gopal2} available for 
${\bar K}N$ scattering for invariant collision energies 
1.48${\le}\sqrt{s_2}{\le}$2.17 GeV. 
For $\sqrt{s_2}{\le}$1.48~GeV  
we adopt the $K$-matrix solution of Martin and Ross~\cite{Martin1}, which
satisfactorily describes all available experimental results below the 
energy of 1.48~GeV. 
The $K$-matrix solution includes only $S$-waves for the 
$I$=0 and $I$=1 channels. However, the data shown in Ref.~\cite{Martin1} 
illustrate that for $\sqrt{s}{\le}$1.48~GeV the contribution from
higher partial waves is small.

\begin{figure}[t]
\vspace*{-4mm}
\centerline{\hspace*{4mm}\psfig{file=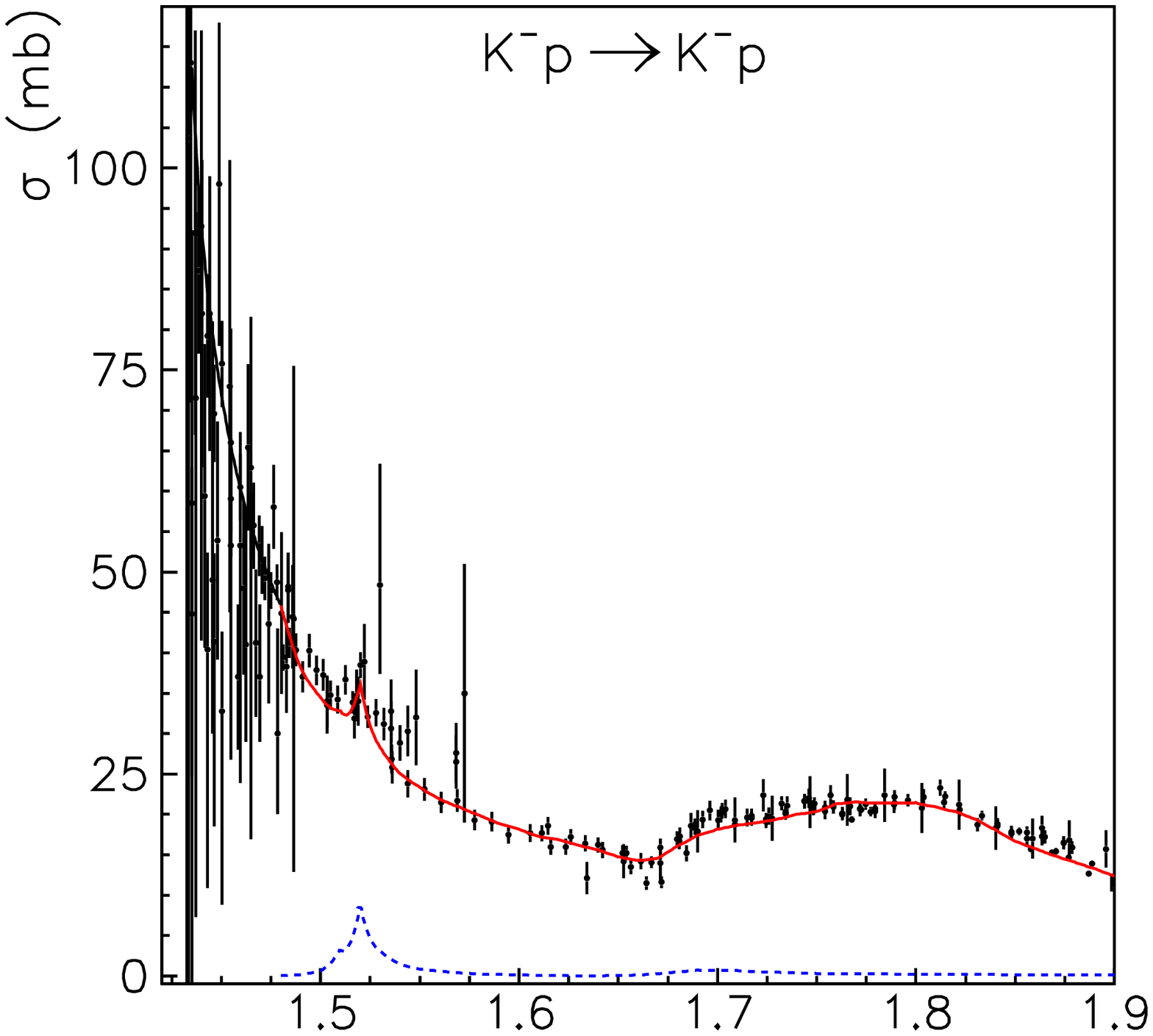,width=9.1cm,height=6.5cm}}
\vspace*{-8mm}
\centerline{\hspace*{4mm}\psfig{file=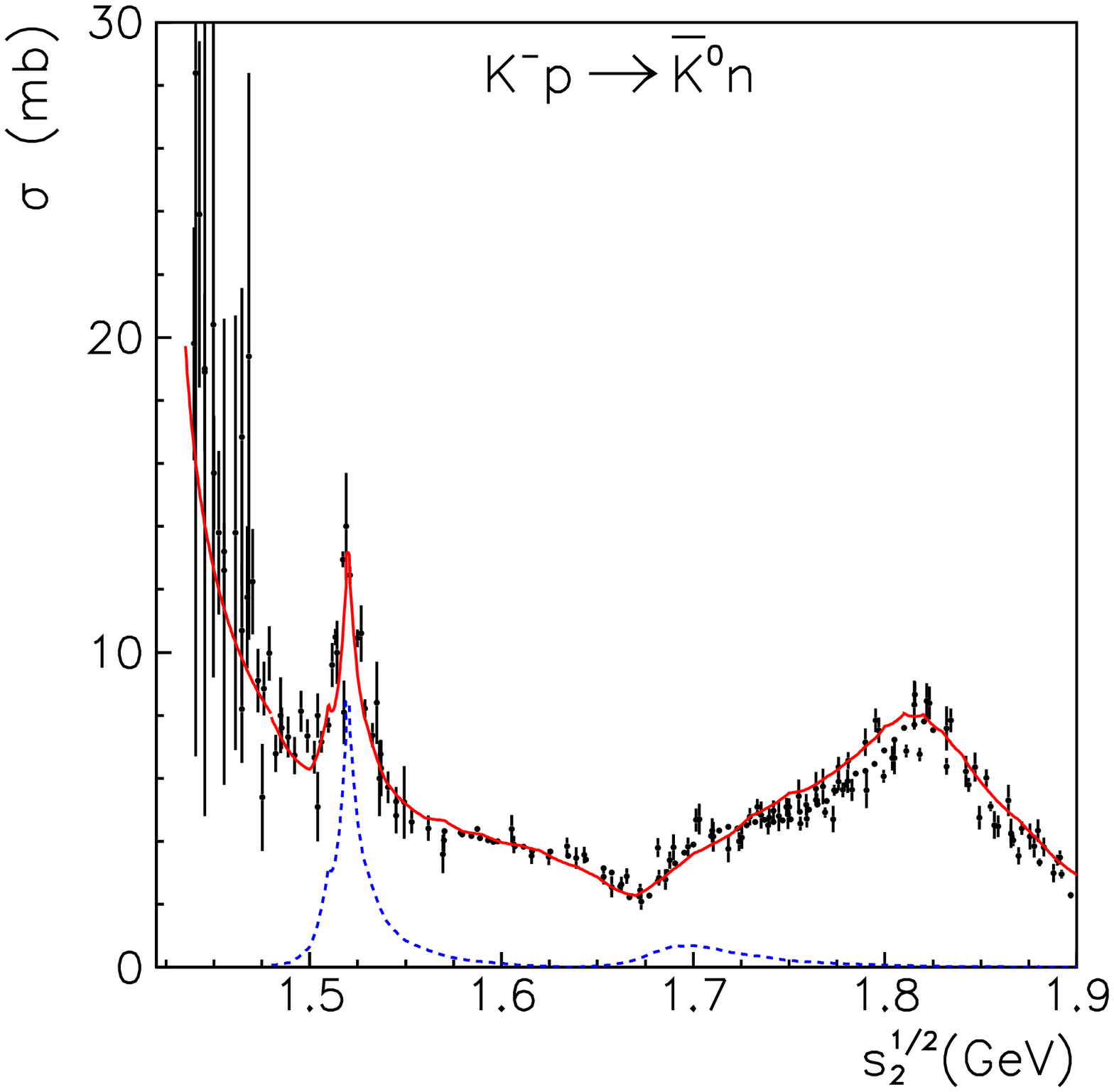,width=9.1cm,height=6.5cm}}
\vspace*{-2mm}
\caption{The $K^-p{\to}K^-p$ and $K^-p{\to}{\bar K^0}n$ cross sections
as a function of the invariant collision energy. 
The solid lines show results based on the 
full reaction amplitude, while the dashed lines indicate contribution from 
the $D_{03}$ partial wave alone.
The data are taken from Refs.~\cite{PDG,Adams,Mast}. 
}
\label{phot8}
\end{figure}

The usefulness of the Drell approach with respect to analyzing the data 
becomes obvious by first looking at results for the cross section of 
the reactions $K^-p{\to}K^-p$ and $K^-p{\to}{\bar K^0}n$, shown in 
Fig.~\ref{phot8}. Here the experimental information, taken from 
Refs.~\cite{PDG,Adams,Mast}, is compared with calculations
utilizing the total $\bar KN$ scattering amplitude 
of Ref.~\cite{Gopal2} (solid lines) with results that take into
account only the contribution from the $D_{03}$ partial wave (dashed lines).
(We use here the standard nomenclature $L_{I \, 2J}$.)
Note that the $D_{03}$ partial wave contains the ($I(J^P)= 0({3 \over 2}^-)$)
$\Lambda$(1520) resonance with nominal mass $1519.5 \pm 1.0$ MeV and full 
width $15.6{\pm}1.0$~MeV~\cite{PDG}. 
While the $K^-p{\to}K^-p$
reaction shows almost no trace of the $\Lambda$(1520), the charge-exchange
reaction $K^-p{\to}{\bar K^0}n$ clearly indicates a resonance structure.
This difference can be understood by recalling that the reaction amplitude for 
the $K^-p{\to}K^-p$ channel consists of (half of) the sum of the $I{=}0$ 
and $I{=}1$ amplitudes, while the $K^-p{\to}{\bar K^0}n$ reaction is given 
by (half of) their difference. 
Since both $I{=}0$ and $I{=}1$ amplitudes contain
large nonresonant contributions, it turns out that this nonresonant
background cancels to a large extent for the $K^-p{\to}{\bar K^0}n$ 
amplitude. Fig.~\ref{phot8} demonstrates quite strikingly that the 
contribution of the $D_{03}$ partial wave to the $K^-p{\to}K^-p$ reaction 
is almost negligible compared with the nonresonant background, 
while its contribution to the charge-exchange channel is sizeable. 

\begin{figure}[b]
\vspace*{-10mm}
\centerline{\hspace*{4mm}\psfig{file=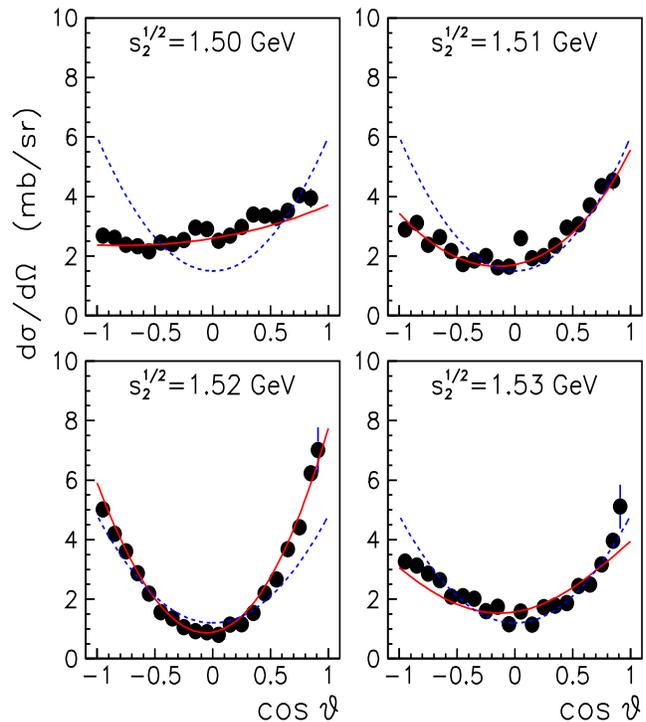,width=9.8cm,height=11.cm}}
\vspace*{-5mm}
\caption{Differential cross section for $K^-p$ scattering 
as a function of the $K^-$ meson scattering angle in the c.m. system 
for different invariant collision energies, $\sqrt{s_2}$. 
The solid lines show results based on the full reaction amplitude, 
while the dashed lines indicate the distribution given
by $1{+}3\cos^2\theta$ normalized to the data.
The data are taken from Refs.~\cite{Adams,Mast}. 
}
\label{phot9}
\end{figure}

This observation suggests that kaon exchange (viz. the Drell mechanism) 
should not produce a pronounced $\Lambda$(1520) signal in the invariant 
mass of the $K^-p$ system for the reaction $\gamma{p}{\to}K^+K^-p$ 
but only in the $\gamma{p}{\to}K^+{\bar K^0}n$ channel. 
Consequently, if the reaction $\gamma{p}{\to}K^+K^-p$ does indeed show a 
substantial effect of the $\Lambda$(1520) resonance, it is a strong
indication that mechanisms other than the Drell mechanism dominate 
the reaction \cite{Barber}. 
 
Fig.~\ref{phot8} illustrates also that any model calculation of
the reaction $\gamma{N}{\to}K{\bar K}N$ where the employed 
${\bar K}N{\to}{\bar K}N$ amplitude is constructed from hyperon 
resonances alone (cf. diagram c) in Fig.~\ref{digdrell})
-- as it is the case in so-called resonance models -- 
is definitely not realistic, since there are significant nonresonant
contributions to the scattering amplitude.  
In this context let us emphasize that 
the ${\bar K}N{\to}{\bar K}N$ amplitude contains large contributions
from higher partial waves. Fig.~\ref{phot9} shows the angular
distribution of the $K^-$ mesons in the c.m. system of the reaction 
$K^-p{\to}K^-p$ reaction, which clearly manifests the increasing importance 
of higher partial waves with increasing invariant collision energy,  
$\sqrt{s_2}$.

With the partial-wave decomposition for the amplitudes $G$ and $H$
given in Eq.~(\ref{pwa}) the differential cross section can be readily 
evaluated for 
the specific $D_{03}$ partial wave where the $\Lambda$(1520) resonance occurs: 
\begin{eqnarray}
\frac{d\sigma}{d\cos\theta} \ = \ &|G|^2+|H|^2 
\ = \ 
\left|\frac{T^-_2(s_2)}{q_2}\right|^2\!\!\!(3\cos^2\theta{+}1) \ .
\end{eqnarray}

The resulting angular dependence, $1{+}3\cos^2\theta$, is shown by
the dashed lines in Fig.~\ref{phot9}, arbitrarily normalized
to the data. It is evident that the $1{+}3\cos^2\theta$ function alone
does not describe the data at $\sqrt{s}$=1520~MeV;
additional contributions, from other partial waves, are required. 
The solid lines in Fig.~\ref{phot9} represent results 
with the total reaction amplitude, i.e. with all partial waves 
including the $D_{03}$. It is interesting to note that 
while the $\Lambda$(1520) resonance remains almost undetectable in the 
$K^-p \to K^-p$ reaction cross section shown in Fig.~\ref{phot8}, because of
the large background, this resonance can be well
reconstructed from an analysis of the angular distributions.
A detailed discussion of the $K^-$ meson angular spectra at energies
around the $\Lambda$(1520) resonance is given in Ref.~\cite{Watson}.

\begin{figure}[t]
\vspace*{-4mm}
\centerline{\hspace*{4mm}\psfig{file=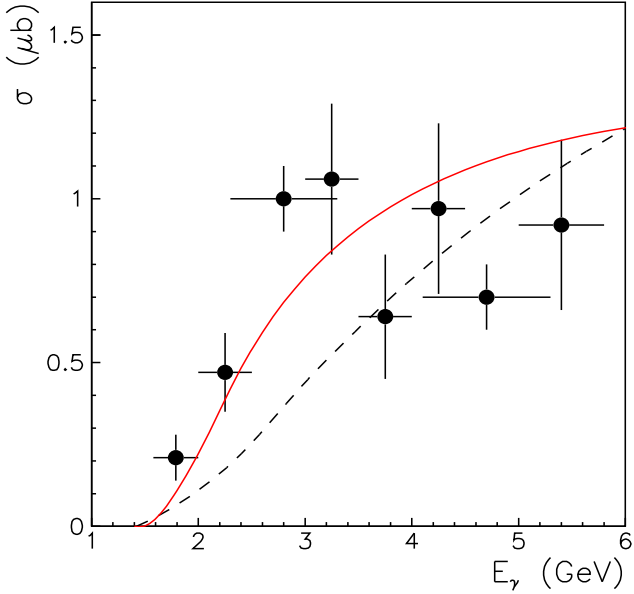,width=9.4cm,height=9.cm}}
\vspace*{-3mm}
\caption{Cross section of the reaction 
$\gamma{p}{\to}K^+K^-p$ as a function of the photon energy. The dashed 
line is the result obtained with a constant reaction amplitude while the
solid line shows the result for the Drell mechanism.
The symbols represent data collected in Ref.~\cite{Baldini}. 
}
\label{phot1}
\end{figure}

Fig.~\ref{phot1} shows the cross section of the 
$\gamma{p}{\to}K^+K^-p$ reaction as a function of photon energy.
The circles represent experimental information collected 
in Ref.~\cite{Baldini}.

The total $\gamma{p}{\to}K^+K^-p$ cross section is
given by integration of the Chew-Low distribution~\cite{Chew}
\begin{eqnarray}
\frac{d\sigma}{ds_2 dt_1} = \frac{|{\mathcal M}_{K^+}{+}{\mathcal M}_{K^-}|^2}
{2^9 \pi^3 (s-m_N^2)^2}\, \frac{\lambda^{1/2}(s_2,m_K^2,m_N^2)}{s_2} \ ,
\end{eqnarray}
where the function $\lambda$ is defined by
\begin{eqnarray}
\lambda (x,y,z)=\frac{(x-y-z)^2-4yz}{4x} \ .
\end{eqnarray}
The dotted line in Fig.~\ref{phot1} is the result with 
a constant invariant $\gamma{p}{\to}K^+K^-p$ reaction amplitude, i.e. 
${\mathcal M}{=}const.$, suitably adjusted to the data 
The full calculation (solid line in Fig.~\ref{phot1}) 
reproduces the energy dependence of the data reasonably
well from threshold up to a photon energy of $\simeq$5~GeV -- 
i.e., over the whole energy range covered by the LAMP2 experiment. 
Also in this case the absolute value of the model result was adjusted 
to the data, namely by tuning the cutoff mass in the form factor 
Eq.~(\ref{ff}), cf. above. Indeed, the experimental $KN$ invariant 
mass spectra which we want to investigate with our model are given 
without absolute normalization, so that the overall normalization 
is irrelevant for our application anyway. 

\section{Invariant {\boldmath $KN$} mass spectra from the 
reaction {\boldmath $\gamma{p}{\to}K^+K^-p$}}

\begin{figure}[b]
\vspace*{-4mm}
\centerline{\hspace*{6mm}\psfig{file=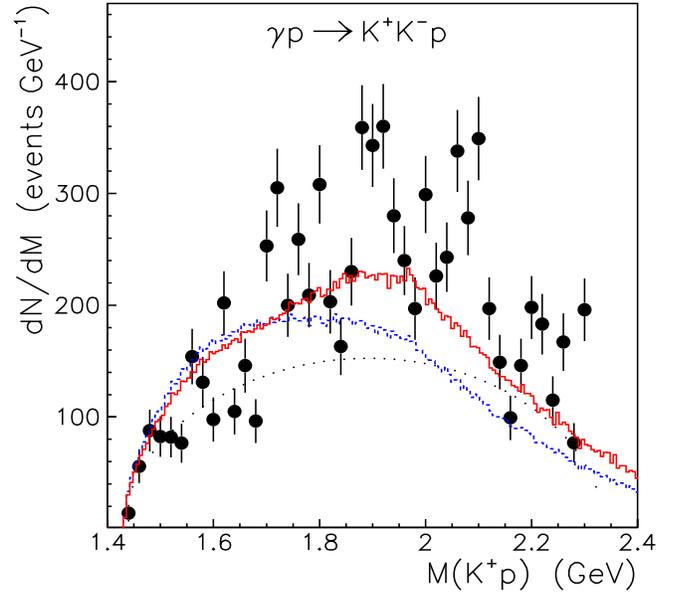,width=9.4cm,height=9.cm}}
\vspace*{-3mm}
\caption{The $K^+p$ invariant mass spectrum from the
$\gamma{p}{\to}K^+K^-p$ reaction. 
The dotted line is the result obtained with a constant reaction amplitude
for the fixed photon energy 3.8 GeV. 
The dashed line shows the result for the Drell mechanism, while the 
solid line is obtained after inclusion of additional $K^\ast$ meson 
exchange for $\Lambda$(1520) production. Here the photon energy is 
averaged over 2.8${<}E_\gamma{<}$4.8~GeV in consistency with the
experiment. The data are taken from Ref.~\cite{Barber}. 
}
\label{phot5}
\end{figure}

\begin{figure}[t]
\vspace*{-4mm}
\centerline{\hspace*{4mm}\psfig{file=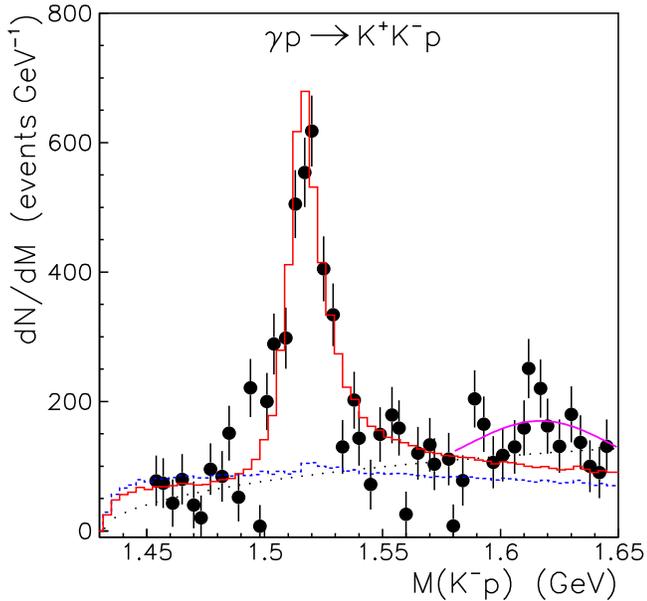,width=9.2cm,height=9.cm}}
\vspace*{-3mm}
\caption{The $K^-p$ invariant mass spectrum for the reaction
$\gamma{p}{\to}K^+K^-p$. Same description of the curves as in
Fig.~\ref{phot5}. 
The thin solid line indicates the contribution of a $\Lambda$(1600)
$P_{01}$ resonance, discussed in Sect. \ref{sec:ELSA}. 
The circles are data taken from Ref.~\cite{Barber}. 
}
\label{phot6}
\end{figure}

The $K^+p$ and $K^-p$ invariant mass spectra were measured by the 
LAMP2 Group~\cite{Barber} in the reaction $\gamma{p}{\to}K^+K^-p$ 
with a tagged photon beam with an energy 2.8${<}E_\gamma{<}$4.8~GeV. 
Figs.~\ref{phot5},\ref{phot6} show the
invariant mass spectra for the $K^+p$ and  $K^-p$ subsystems.
Note that for the photon energy of 4.8~GeV the maximal invariant 
mass of the $KN$ system ranges up to roughly 2.65~GeV. The
experimental $K^-p$ mass spectrum is provided only up to $\simeq$1.65~GeV
in Ref.~\cite{Barber}, while the data for the $K^+p$ mass distribution are 
given over almost the whole available range.

The dotted lines in Figs.~\ref{phot5},\ref{phot6}
represent the phase space distribution 
\begin{eqnarray}
\frac{d\sigma}{ds_2}{=}\frac{\lambda^{1/2}(s,s_2,m_K^2)
\lambda^{1/2}(s_2,m_N^2,m_K^2)}{2^8\pi^3\,s\,s_2\,(s-m_N^2)}\,
|{\mathcal M}|^2,
\label{phase}
\end{eqnarray}
evaluated for the fixed averaged photon energy of 3.8~GeV and 
assuming a constant reaction amplitude, i.e. ${\mathcal M}{=}const.$ In our
notation the invariant $KN$ mass $M(KN)$ is given by $M(KN){=}\sqrt{s_2}$.
Since the experimental results are available only with arbitrary normalization 
we adjust the amplitude ${\mathcal M}$ to the $K^-p$ background 
at invariant masses above the resonance structure corresponding to the 
$\Lambda(1520)$, i.e. around 1570 MeV. The same normalization factor is
then used also in the evaluation of the $K^+p$ mass spectrum.

{}From Fig.~\ref{phot5} it is obvious that the phase space distribution 
differs from the measured mass dependence of the $K^+p$ spectrum. 
On the other hand the measurement exhibits large fluctuations and 
therefore it is hard to say to what extent the structure of the data 
reflects underlying physics or whether it is simply a consequence of
poor statistics. 
Nonetheless, it appears that there is 
an excess of events at high $K^+p$ invariant masses. 
Note that the incident photon
energy distribution in the experiment is proportional to $E_\gamma^{-1}$
and, therefore, naively one expects more events at low $K^+p$ masses. 
Fortunately, the $K^-p$ data themselves indicate a possible 
explanation of that 
problem, namely the presence of a 
clean resonance structure around $M(K^-p)$ =1520~MeV and probably at around
$M(K^-p)$=1620~MeV, cf. Fig.~\ref{phot6}. 
One would expect that a large contribution at low $K^-p$ masses, like that  
associated with the $\Lambda$(1520) resonance,
would be kinematically reflected in the $K^+p$ mass spectrum 
and cause a substantial shift to higher $M(K^+p)$ values in the latter.

Within an analysis based on the 
Drell mechanism the resonances at $M(K^-p)$ = 1520~MeV and 
$M(K^-p)$=1620~MeV enter via the elastic $K^-p{\to}K^-p$
scattering amplitude. Indeed the partial wave analysis includes
the $\Lambda$(1520) $J^P{=}\frac{3}{2}^-$ 
and $\Lambda$(1600) $J^P{=}\frac{1}{2}^+$ resonances 
in the $D_{03}$ and the $ P_{01}$ partial waves, respectively.
However, the elastic scattering data depicted in Fig.~\ref{phot8} 
as well as the $K^-p$ amplitude of the PWA do not show any obvious 
signal of those resonances, as already pointed out earlier. 
Therefore, we do not expect that a calculation 
based on the Drell mechanism will be able to reproduce these
resonances as observed in the $K^-p$ invariant mass spectrum
of the reaction $\gamma{p}{\to}K^+K^-p$. 

This is indeed the case as can be seen from 
our results for the Drell diagrams which correspond to 
the dashed histograms in Figs.~\ref{phot5} and \ref{phot6}.
Obviously, there is 
only a small, hardly noticeable enhancement in the $K^-p$ mass spectrum 
due to the $\Lambda$(1520) resonance and the same is the case also for the
$\Lambda$(1600) resonance. The $E_\gamma^{-1}$ factor of the
photon energy distribution is now included in the calculation and it is
obvious from Fig.~\ref{phot5} how the spectrum is shifted to lower
$K^+p$ invariant masses because of that. Note that the results were
normalized in such a way that we are in line with the low invariant-mass
spectrum of the $K^-p$ system and we use again the same normalization
for the $K^-p$ and $K^+p$ results. Also, 
our calculations were developed in line with the S\"oding model~\cite{Soding1} 
in order to allow for possible kinematic cuts and simulation of the 
detector acceptance of current experiments. The integration over phase 
space is based on the Monte-Carlo method that allows one to construct an
event generator. That is why the calculations are shown as 
histograms.

The results in Fig. \ref{phot6} make it clear 
that additional reaction mechanisms need to be considered if one wants
to describe the experimental $K^-p$ invariant mass spectrum. 
To account for such additional mechanisms we resort here 
to $K^\ast$ meson exchange with subsequent excitation 
of the $\Lambda$(1520) resonance, as depicted by the diagram d) in 
Fig.~\ref{digdrell}. 
Indeed, the experimental helicity-frame angular distribution of the 
$\Lambda$(1520) decay for the reaction $\gamma{p}{\to}K^+K^-p$ 
\cite{Barber} strongly suggests that $K^*$ exchange should play
an important role. 
Note, however, that 
while our calculation of the Drell mechanism is considerably constrained by 
experimental information on the $KN$ and $\bar KN$ amplitude and 
therefore can be considered as a solid background for the reaction
$\gamma{N}{\to}K{\bar K}N$, 
this is not the case for the $K^\ast$ exchange because there are no
data on the reaction $K^\ast N\to KN$ (or $\overline{K^\ast} N\to \bar KN$).
Furthermore, it is possible that many different
$\Lambda$ and $\Sigma$ hyperon resonances are excited in the
$\gamma{N}{\to}K{\bar K}N$ reaction through $K^\star$ meson
exchange. But the relevant coupling constants of the $K^\ast$ meson
to the hyperon resonances are completely unknown. Thus, the best  
one can do is to start with the general structure of the 
$K^\star$ exchange amplitude for hyperon resonance photoproduction 
and to fit the unknown parameters to the presently available data. 
For reasons of simplicity we decided to take into account only the 
excitation of the $\Lambda$(1520) resonance. This has the advantage
that one can at least use the available data on the reaction
$\gamma p \to K^+\Lambda$(1520) to constrain the parameters.
Details about the structure of the employed $K^\star$ exchange 
amplitude are given in the Appendix, together with a 
comparison to the $\gamma{p}{\to}K^+\Lambda$(1520) data.

We should also mention that, of course, it is only an assumption that
the needed additional contributions come from $K^\star$ exchange 
alone. In principle, any other meson exchange could contribute as well.  
And even scenarios like those considered in Refs.~\cite{Nam1,Nam2}, 
where the bulk of the $\gamma p \to K^+\Lambda$(1520) cross section
is generated by contact terms, are possible. 
 
The solid histograms in Figs.~\ref{phot5} and \ref{phot6} represent
the $K^+p$ and $K^-p$ invariant mass spectra evaluated with
$K$ and $K^\ast$ meson exchanges. 
The calculation reproduces quite well the experimental 
$K^-p$ invariant mass spectra. Note that we do not include 
the $\Lambda$(1600) resonance excitation via $K^\ast$ meson 
exchange since there is no
experimental information about the energy and $t$-dependence
of its production and also no data on the spin density matrix. 
As a consequence our calculation underestimates the 
$K^-p$ spectrum in the corresponding invariant mass region. 
However, we will come back to this issue in the next section.
 
The inclusion of $K^\ast$ meson exchange also improves the
description of the $K^+p$ invariant mass spectrum. In particular, 
we now find a noticeable enhancement at higher invariant masses. 
However, we still underestimate the $K^+p$ mass spectrum at masses
above 1.8~GeV. The discrepancy might be related to the
production of other hyperon resonances with masses above 
1.650~GeV, i.e. in the $K^-p$ invariant mass region not 
covered by the LAMP2 experiment \cite{Barber}. If such 
resonances can be produced by $K^\ast$ meson exchange they would 
be seen in the $K^-p$ invariant mass distribution -- and adding 
their contributions to the $\gamma{p}{\to}K^+K^-p$ amplitude 
might improve the description of the $K^+p$ mass spectrum. 
However, at this stage and without data, this remains pure speculation.

%%%%%%%%%%%%%%%%%%%%%%%%%%%%%%%%%%%%%%%%%%%%%%%%%%%%%%%%%%%%%%%%%%%%%%%%%%%
\section{Application to the SAPHIR data}
\label{sec:ELSA}
As a first application of our model and in order to demonstrate
its potential for future analyses we present here a comparison
with data on the reaction $\gamma{N}{\to}K{\bar K}N$ taken recently 
by the SAPHIR collaboration at ELSA (Bonn). Their study of the channel
$\gamma{p}{\to}K^+K^0_sn$ was among the first which revealed
evidence for the $\Theta^+$(1540) pentaquark~\cite{Barth}. However, the group
has also provided invariant mass spectra for the $K^-p$ and $K^+p$
systems from the reaction $\gamma{p}{\to}K^+ K^-p$~\cite{Barth,SAP2}
which we want to analyze now. In their experiment the photon
energy range is $1.74{\le}E_\gamma{\le}2.6$~GeV, i.e. significantly
lower than in the LAMP2 experiment. 

\begin{figure}[t]
\vspace*{-4mm}
\centerline{\hspace*{4mm}\psfig{file=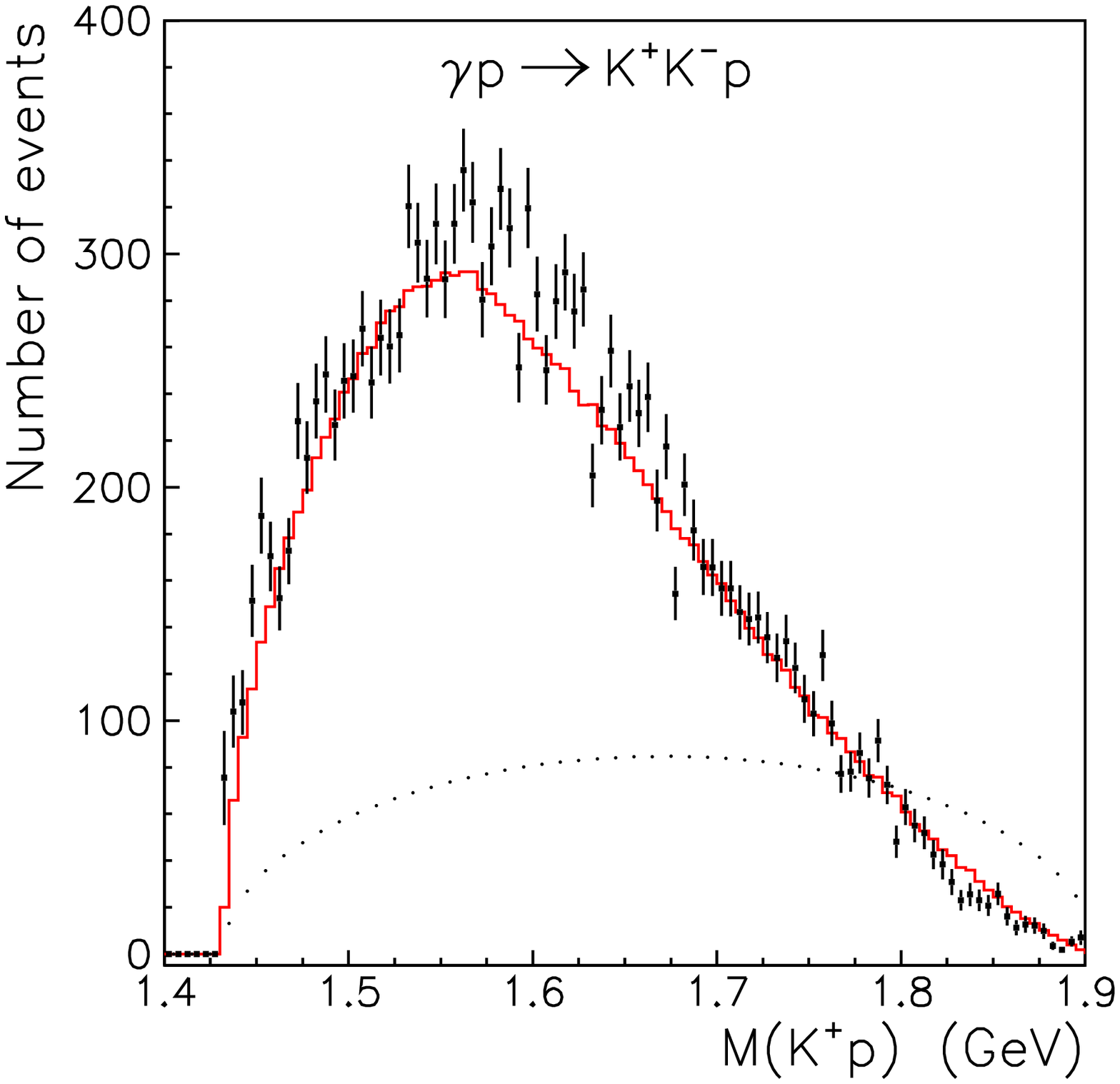,width=9.2cm,height=9.cm}}
\vspace*{-10mm}\centerline{\hspace*{4mm}\psfig{file=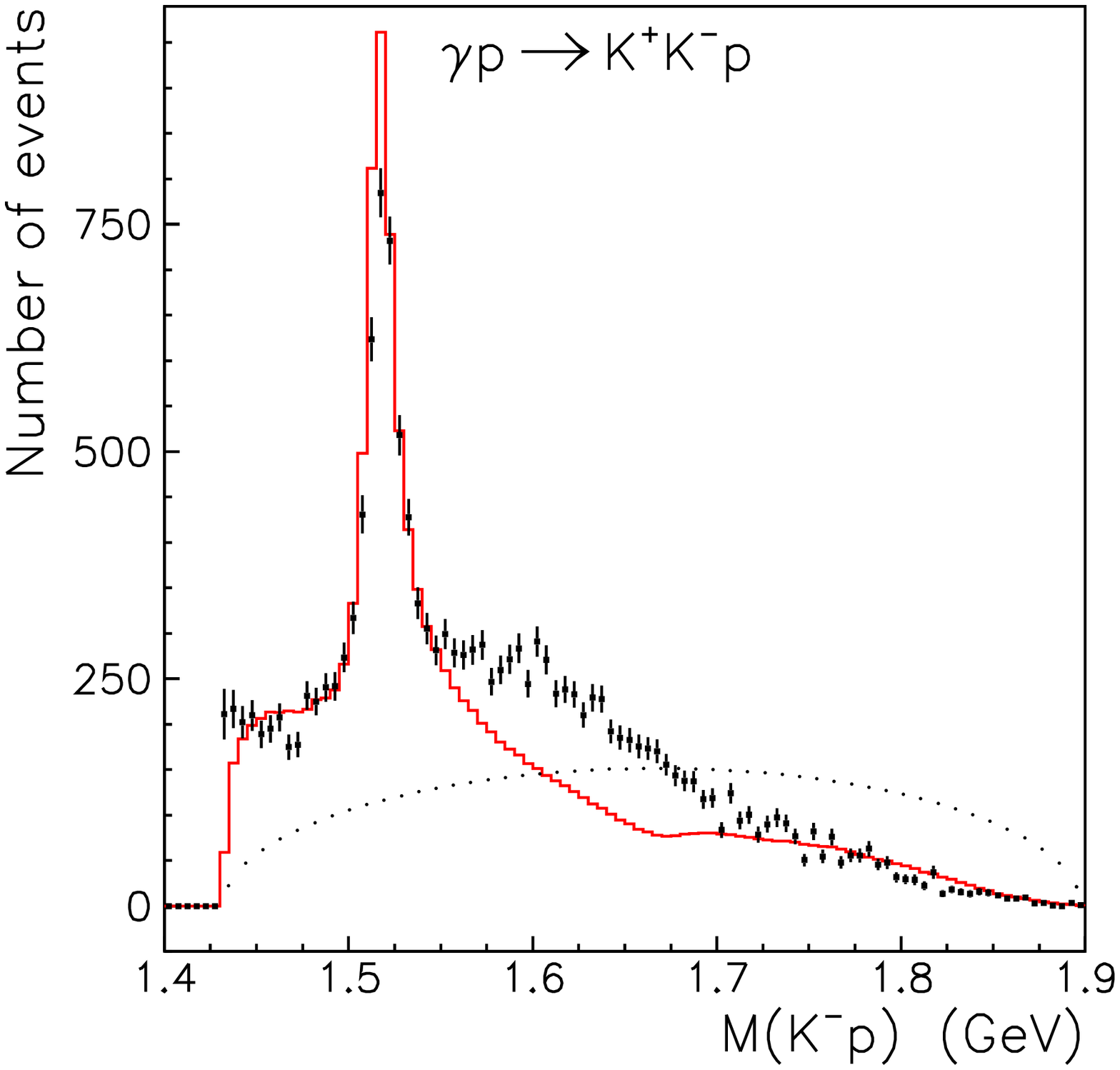,width=9.2cm,height=9.cm}}
\vspace*{-3mm}
\caption{The $K^+p$ and $K^-p$  spectra for the
$\gamma{p}{\to}K^+K^-p$ reaction measured by SAPHIR 
Collaboration~\cite{Barth,SAP2}. Same description of curves 
as in Fig. \ref{phot6}.}
\label{phot3b}
\end{figure}

\begin{figure}[t]
\vspace*{-4mm}
\centerline{\hspace*{4mm}\psfig{file=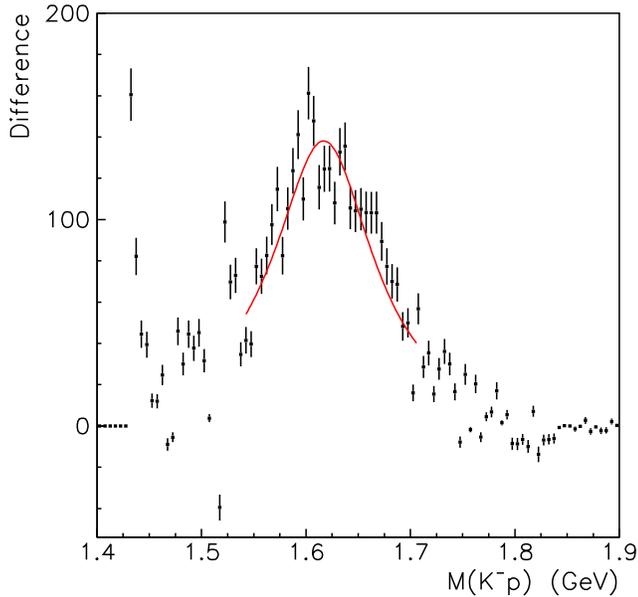,width=9.2cm,height=9.cm}}
\vspace*{-4mm}
\caption{The difference between the SAPHIR data~\cite{Barth,SAP2}
for the $K^-p$ invariant mass spectra of $\gamma{p}{\to}K^+K^-p$
and our model prediction. The solid line indicates the fit with a
relativistic Breit-Wigner function.}
\label{phot3c}
\end{figure}

The results of our model (including both the Drell mechanism and the
$K^*$ exchange contribution) are shown in Fig.~\ref{phot3b} 
for the SAPHIR experiment. Since the data
are without absolute normalization we readjusted our model 
predictions to the maximum of the $K^+p$ invariant mass spectrum. 
It is obvious from Fig.~\ref{phot3b} that the $K^+p$ mass spectrum
is nicely reproduced over the whole invariant-mass range. 
Moreover, at the same time (and with the same normalization) 
there is also excellent agreement with the $K^-p$ spectrum 
in the region of the $\Lambda$(1520) resonance, as well as at
high invariant masses, cf. Fig.~\ref{phot3b}. However, there is 
some excess in the experimental mass spectrum around 1600 MeV. 
If one subtracts the model prediction from
the SAPHIR data one obtains the results shown in Fig.~\ref{phot3c}.
It is interesting to see that this difference strongly resembles 
a resonance signal, something one would not have guessed easily 
from the original data in Fig.~\ref{phot3b}. Fitting this difference
with a relativistic Breit-Wigner amplitude Eq.~(\ref{Wigner}) 
yields the values
\begin{equation}\label{Rpara}
M_R = 1617 \pm 2 ~{\rm MeV}\, , \quad \Gamma_R  = 117\pm 4 ~{\rm MeV}\, ,
\end{equation}
with a $\chi^2$ per data point of $\chi^2/N$=4. The large $\chi^2/N$ reflects
the considerable fluctuation of the data.
 
We are inclined to identify this structure with the 
$\Lambda$(1600) $P_{01}$ resonance \cite{PDG}, though we are aware 
that further and more careful analyses are needed in order to
substantiate such a claim, specifically because the PDG lists also
other hyperon resonances in this energy region \cite{PDG}, e.g.
the $\Sigma$(1580) and $\Sigma$(1620).
Independently of that, we believe that this particular case already
demonstrates very clearly the power of a reliable
model for the background to the reaction $\gamma{N}{\to}K{\bar K}N$ 
in the analysis of experimental data. Two further short comments
are in order: First, the SAPHIR collaboration is in the process of
analyzing the $\Lambda (1520)$ signal in their data~\cite{SAP3} and,
second, we speculate that the sharp signal on the left side of 
Fig.~\ref{phot3c} is related to the excitation of the $\Lambda (1405)$
through $K^*$ exchange. This deserves further study.

As a consistency check, we have added the contribution of the $\Lambda$(1600)
resonance to the $K^-p$ invariant mass spectrum of Ref.~\cite{Barber},
as shown by the thin solid line in Fig.~\ref{phot6}. One can see that its 
contribution is indeed consistent with these data, however, the figure
also underlines our earlier statement that the data from \cite{Barber}
are not sufficiently precise to allow an extraction of resonance properties for 
invariant masses above the $\Lambda$(1520). 

Coming back to the $K^+p$ mass spectrum of the SAPHIR collaboration 
we would like to emphasize that its dependence on the invariant
mass $M(K^+p)$ is rather smooth up to the highest values
as can be seen in Fig.~\ref{phot3b}. Therefore, we believe 
that the structures seen in the corresponding LAMP2 data,
cf. Fig.~\ref{phot5}, are most likely  
fluctuations associated with low statistics. 
This concerns presumably also events at those higher 
invariant masses which are not covered by the SAPHIR data. Anyway,
it would be interesting to explore this region again 
experimentally with much better statistics than what was available
in the LAMP2 experiment. 

\begin{figure}[t]
\vspace*{0mm}
\centerline{\hspace*{1mm}\psfig{file=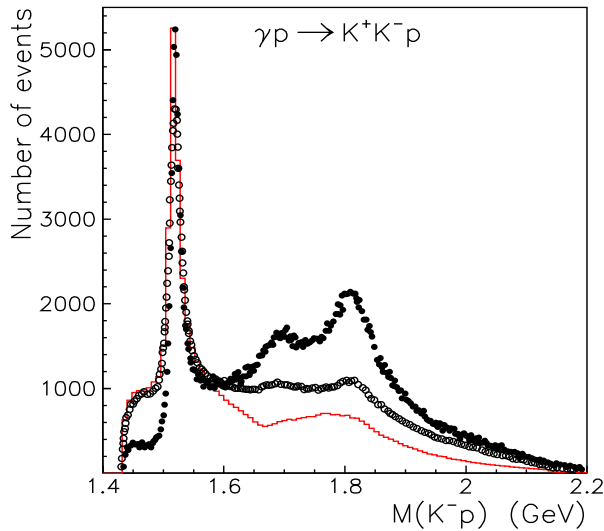,width=8.7cm,height=8cm}}
\vspace*{-2mm}
\caption{The $K^-p$ spectrum for the reaction $\gamma{p}{\to}K^+K^-p$ as
measured by the CLAS Collaboration~\cite{Kubarovsky1}. The open and closed
circles show results obtained with different criteria for the
final-state particle reconstruction. The data are not efficiency corrected.
The solid histogram is our calculation for the photon energies
$1.8{\le}E_\gamma{\le}3.8$ GeV.
}
\label{phocla3a}
\end{figure}

\section{Application to the CLAS data}
\label{sec:CLASs}

Recently also the CLAS Collaboration reported~\cite{Kubarovsky1} new
results on the reaction $\gamma{p}{\to}K^+K^-p$ for photon energies of
$1.8\le E_\gamma{\le}3.8$ GeV. However, unfortunately their data
are not efficiency corrected~\cite{Vita}. Thus a direct comparision
of our results with those data and specifically, an analysis
similar to the one for the SAPHIR data with the aim to
extract possible contributions from hyperon resonances is not
possible at this stage. Still we find it interesting to compare the
prediction for the $K^-p$ invariant mass spectrum resulting from the
Drell mechanism with those data.

The solid histogram in Fig.~\ref{phocla3a} shows
the result of our calculation based on the diagrams a) and b) of
Fig.~\ref{digdrell} and taking into account additional
contributions to the $\Lambda{(1520)}$ production due to
$K^\ast$-exchange. The uncorrected experimental spectra are from
Ref.~\cite{Kubarovsky1}, where the two
sets of data points (open and closed squares) were
obtained by using different criteria for the final-state particle
reconstruction. For the comparision we normalized both sets of the data
and our calculation at the maximal yield of the $\Lambda$(1520) resonance.
Our calculation
clearly indicates that the $K^-p$ mass spectra at
$1.8{\le}E_\gamma{\le}3.8$ GeV provide an excellent tool to
identify possible additional contributions from high-mass hyperon
resonances.
Also, an additional measurement of the angular spectra in the
Gottfried-Jackson system at fixed $K^-p$ masses would provide valuable
information for the hyperon resonances spectroscopy.
Thus, it will be interesting to analyze the CLAS data within our model
once these data are efficiency corrected. Only then concrete
conclusion can be made.

\section{Conclusion}
We have studied the reaction $\gamma{N}{\to}K{\bar K}N$ utilizing the
Drell mechanism and taking into account the full reaction amplitudes 
for the $KN$ and $\bar KN$ subsystems. 
Our results show that the Drell mechanism for kaon exchange 
alone is not sufficient to describe the available data on the $K^+p$ and 
$K^-p$ invariant mass spectra. Thus, we have included $K^\ast$ meson 
exchange as an additional reaction mechanism. By assuming a large $K^\ast$ 
coupling to the $\Lambda$(1520) resonance a quantitative description
of the $K^-p$ invariant mass spectrum can be achieved. Moreover, 
our model calculation also yields a good overall reproduction of the 
cross section data for the photoproduction of the 
$\Lambda$(1520) resonance, i.e of its energy and $t$-dependence.

In the paper by Barber et al. \cite{Barber} it was argued, based
on the measured distribution of the $\Lambda$(1520) decay into the $K^-p$
channel in the Gottfried-Jackson frame, that 
kaon exchange alone cannot account for the $\Lambda$(1520) 
production mechanism. Our investigation, utilizing the Drell mechanism and 
experimental information on the $KN$ and $\bar KN$ scattering 
amplitudes, can be considered as an independent confirmation of
this conclusion. 

The $\Lambda$(1520) excitation in photoproduction illustrates an aspect 
that is also relevant for discussions concerning the $\Theta^+$ pentaquark.
While the $\Lambda$(1520) resonance remains practically undetectable 
in $K^-p$ elastic scattering (see Fig.~\ref{phot8}), 
the resonance is clearly visible in the photoproduction reaction shown in 
Fig.~\ref{phot6}. Evidently, the difference between elastic scattering 
and photoproduction must involve reaction mechanisms or channels which 
are not accessible in elastic scattering -- 
such as the $K^\ast$ meson exchange which we assumed in the present study. 
The situation for $\Theta^+$(1540) production might be similar. Different 
reaction mechanisms/channels could be quite selective to the $\Theta^+$ 
excitation and, consequently, could be a 
natural reason why the pentaquark was observed in some experiments 
but not in others. The $\Lambda$(1520) photoproduction is an 
excellent example of such a situation. 
Clearly, at present, in either case the reaction mechanisms governing the
photoproduction are not yet identified and, moreover, there are no reasons 
to believe that they might be similar, let alone the same. 

As a first application, we have analyzed the SAPHIR data on $\gamma p 
\to p K^+ K^-$ and shown that our approach gives an excellent 
description of the  
$K^+p$ mass distribution for all energies and the $K^-p$ mass distribution
in the vicinity of the $\Lambda (1520)$ resonance 
and for energies above 1.7~GeV.
We have shown that the remaining strength can be well described by a 
resonance with the parameters given in Eq.~(\ref{Rpara}) - this state 
could be the $\Lambda (1600)$ $P_{01}$ resonance of the PDG listing.
 
Since our model allows us to study not only the invariant mass spectra
but also momentum and angular distributions of the final particles, it
is also possible to compare the model results with more exclusive
(i.e. differential) data and to investigate the possible role of the 
detector acceptance and kinematical cuts in the extraction of 
specific resonances, like the $\Theta^+$(1540) pentaquark, 
from present or future experiments~\cite{Hicks}.  

\subsection*{Acknowledgments}
We would like to thank D. Diakonov, K. Hicks, N.N. Nikolaev, W. Sch\"afer and 
S. Stepanyan for useful discussions. We are grateful to J.~Barth, F.~Klein,
M.~Ostrick and W.~Schwille for supplying us with the SAPHIR data and for
comments. This work was partially 
supported by the Department of
Energy under contract DE-AC05-84ER40150 under which SURA operates
Jefferson Lab and by Deutsche Forschungsgemeinschaft through funds
provided to the SFB/TR 16 ``Subnuclear Structure of Matter''. 
This research is part of the EU Integrated Infrastructure Initiative Hadron
Physics Project under contract number
RII3-CT-2004-506078. A.S. acknowledges support by the COSY FFE grant
No. 41445400 (COSY-067).

%%%%%%%%%%%%%%%%%%%%%%%%%%%%%%%%%%%%%%%%%%%%%%%%%%%%%%%%%%%%%%%%%%%%%%%
\appendix
\section{The {\boldmath $K^*$} meson exchange contribution}

In this appendix we specify the $K^*$ meson exchange contribution
to the reaction $\gamma{p}{\to}K^+K^-p$ used in our model
calculation. Since the energy range where the experimental 
information is available, $2.8{\le}E_\gamma{\le}4.8$~GeV, is already
too high for employing effective Lagrangians \cite{Titov8}, we resort to
Regge phenomenology. The reaction mechanism is 
depicted in Fig.~\ref{digdrell}d), where we take into account only
the excitation of the $\Lambda$(1520) resonance. This has the
advantage that we can use available data on the reaction 
$\gamma{p}{\to}K^+\Lambda$(1520) to constrain the 
{}free parameters. 
Clearly, other hyperon resonances could contribute as well and there
should be also a nonresonant background for $K^*$ exchange. 
However, it is impossible to constrain those contributions from
data and therefore we leave them out altogether. 

Let us first specify the amplitude for the elementary 
reaction $\gamma{p}{\to}K^+\Lambda$(1520). 
We use phenomenological helicity amplitudes for the 
single $t$-channel meson exchange Regge pole for the 
process $ab{\to}12$,
\begin{eqnarray}
M^{\lambda_1\lambda_a}_{\lambda_2\lambda_b}(s,t_1)=
-V_x^{\lambda_1\lambda_a}(t_1) R_x(s,t_1) V^x_{\lambda_2\lambda_b}(t_1) \ .
\label{hel}
\end{eqnarray}
Here $R_x$ is the Regge propagator for the exchange of the meson $x$,
\begin{eqnarray}
R_x(s,t_1)=\frac{1{+}(-1)^{s_x}\exp[-i\pi\alpha_x(t_1)]}
{2\,\sin[\pi\alpha_x(t_1)]\,
\Gamma [l_x{-}\alpha(t_1)]}\left[\alpha^\prime s\right]^{\alpha_x(t_1)} \ ,
\label{prop1}
\end{eqnarray} 
where $s_x$ and $l_x$ are the spin of the exchanged meson and the spin of the 
lowest state of the Regge trajectory, respectively. For $K^\ast$ meson exchange
we use $s_{K^*}{=}1$, $l_{K^*}$=1. The Regge trajectory $\alpha_{K^*}$ is taken as
\begin{eqnarray}
\alpha_{K^\ast}=1+\alpha^\prime(t-m_{K^\ast}^2),
\end{eqnarray}
with $\alpha^\prime{=}0.9$~GeV$^{-2}$ and $m_{K^\ast}$ = 892 MeV. 
The vertex function $V$ is parameterized as
\begin{eqnarray}
V_x^{\lambda_1\lambda_a}(t_1)=\beta_x^{\lambda_1\lambda_a}
[t_1-t_{min}]^{|\lambda_1{-}\lambda_a|/2},
\label{flip}
\end{eqnarray}
where $\beta_x$ is a helicity coupling constant and $t_{min}$ is the minimal
four-momentum transfer. The last term in Eq.~(\ref{flip}) ensures that
the spin flip amplitude vanishes at forward direction. 
In principle each vertex should be dressed with a form factor,
which can be determined from the $t$-dependence of the experimental
data. When the masses of the initial and final particles are 
different, i.e. $m_1{\neq}m_a$ and $m_2{\ne}m_b$ then
\begin{eqnarray}
t_{\min}\simeq - \frac{(m_a^2-m_1^2)(m_b^2-m_2^2)}{s}.
\end{eqnarray}
 
\begin{figure}[b]
\vspace*{-5mm}
\centerline{\hspace*{4mm}\psfig{file=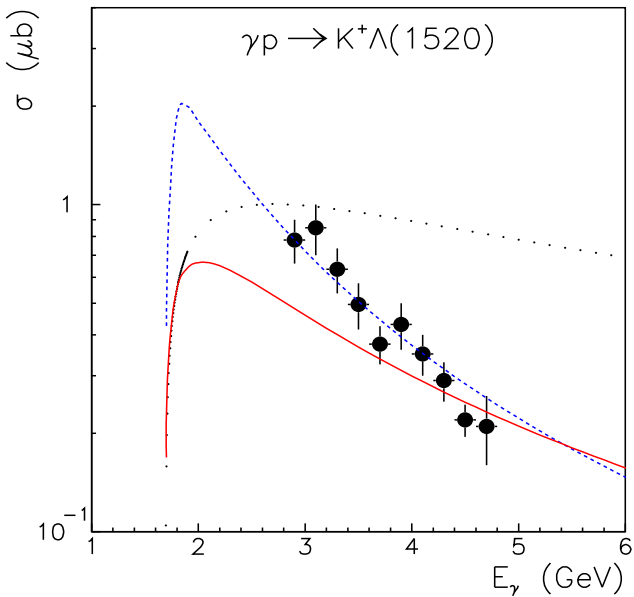,width=9.4cm,height=9.cm}}
\vspace*{-3mm}
\caption{The cross section for the reaction $\gamma{p}{\to}K^+\Lambda$(1520).
The dotted line is the result obtained with constant reaction amplitude. 
The dashed line shows the calculation with $K^\ast$ meson exchange
without form factor and the solid line is the result obtained with
form factor.
The solid circles show data from Ref.~\cite{Barber}. 
}
\label{phot1a}
\end{figure}

Finally, the differential cross section for the reaction
$ab{\to}12$ is given as
\begin{eqnarray}
\frac{d\sigma}{dt}{=}\frac{1}{(2s_a{+}1)(2s_b{+}1)64\pi 
\lambda(s,m_a^2,m_b^2)} \!\sum_{\lambda_i}
|M^{\lambda_1\lambda_a}_{\lambda_2\lambda_b}|^2,
\end{eqnarray}
where $s_a$ and $s_b$ are the spins of the initial particles
and the summation is over all helicity amplitudes. Within 
this approach it is straightforward to include any meson exchange 
and any initial and final states. A detailed comparison
between model calculations and data from pion and kaon induced 
reactions are given in Refs.~\cite{Irving,Kaidalov,Zayats,Szczepaniak}. 

The principal uncertainties of this model are due to the unknown helicity 
couplings and possible contributions from different exchange trajectories.
The model parameters can be fixed only by 
comparison with available experimental results.
For instance, the energy dependence of the reaction cross section 
is driven by the energy dependence of the Regge propagator, which 
can be used for fixing the exchange trajectories.  
The $\gamma{p}{\to}K^+\Lambda$(1520) reaction cross section was 
measured~\cite{Barber,Boyarski} and is shown in Fig.~\ref{phot1a}
as a function of the photon energy. The dotted line shows the
cross section for $M^{\lambda_1\lambda_a}_{\lambda_2\lambda_b}{=}const.$

For $K^\ast$ exchange the $K^\ast{K}\gamma$ coupling constant 
can be determined
from the $K^\ast{\to}K\gamma$ decay, but the couplings 
and possible form factor at the $K^\ast N\Lambda$(1520) vertex 
are unknown. These uncertainties allow enough freedom to reproduce the 
absolute normalization of the data. Also, the energy dependence
of the reaction cross section, and the $t$-dependence of the differential
spectra and density matrix elements can be fitted simultaneously
resulting in a quite reasonable determination of the free
parameters.

\begin{figure}[t]
\vspace*{-4mm}
\centerline{\hspace*{4mm}\psfig{file=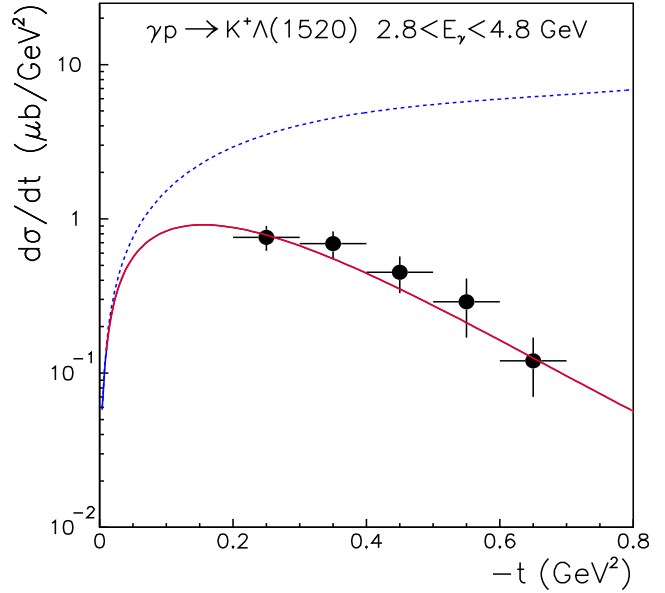,width=9.4cm,height=9.0cm}}
\vspace*{-3mm}
\caption{The $\gamma{p}{\to}K^+\Lambda$(1520) differential cross
section as a function of $-t$. The curves are results of model 
calculations with $K^*$ meson exchange without form factor (dashed)
and with form factor (solid).
The data are from Ref.~\cite{Barber}. 
}
\label{phot11}
%\vspace*{4mm}
\end{figure}

The dashed line in Fig.~\ref{phot1a} shows the result from $K^\ast$ meson 
exchange. It is in line with the data over the photon energy range
covered by the LAMP2 experiment. In this particular fit to the data we 
omitted form factors at the vertices but adjusted instead the overall 
normalization of the total amplitude. Let us mention though, that in principle
the coupling constants for each helicity amplitude 
should be different~\cite{Irving}. 

Additional information on the reaction mechanism is 
contained in the $t$-de{\-}pen{\-}dence. 
Note that the vertex functions of the
helicity amplitudes given by  Eq.~(\ref{flip}) are proportional 
to $(-t)^{n/2}$ with 
$n{=}|\lambda_1{-}\lambda_a|{+}|\lambda_2{-}\lambda_b|$ 
being the net helicity flip. This implies that the spin flip amplitudes
vanish at forward direction and increase with $|t|$. 
Data on the 
differential cross section for $\gamma{p}{\to}K^+\Lambda$(1520) 
as a function of $t$, the square of the four-momentum transferred, 
for photon energies 2.8$<E_\gamma < 4.8$~GeV~\cite{Barber}  
are presented in Fig.~\ref{phot11}. The dashed line is our result 
without form factors. Apparently it is in disagreement with the data.
Indeed the calculated differential cross section increases with
$|t|$, which is an unphysical dependence exhibited by the low $t$ 
Regge model. Thus, 
to describe the data one needs to introduce form factors. We take
those form factors to be $F(t)= \rm{exp}(3t)$ for all helicity amplitudes. 
Furthermore we assume 
that the form factors do not depend on the photon energy. This
assumption is not necessarily correct because other higher-mass $e$ meson 
exchanges might contribute and could be dressed with different 
form factors. Alternatively one can introduce absorptive 
corrections~\cite{Irving,Gottfried2,Levy} that allow one to reproduce
the $t$-dependence phenomenologically.

The solid lines in Figs.~\ref{phot1a} and~\ref{phot11} show results
for $\gamma{p}\to K^+\Lambda$(1520) calculated with the inclusion of the
form factor $F(t)=\rm{exp}(3t)$ for each helicity amplitude 
in Eq.~(\ref{hel}). 
Now the differential cross section is well reproduced by the model. 
However, at the same time the description of the integrated 
cross sections deteriorates somewhat. In order to achieve a 
better overall description of the experiment one could 
either readjust the $K^\ast$ meson exchange trajectory or 
assume that the form factor depends also on the energy. The latter case
is more natural since the data generally indicate that the slope of the 
$t$-dependence depends on energy. Experimental results on the 
energy dependence of the slope of the $t$-dependence for different 
photoproduction reactions are reviewed in 
Refs.~\cite{Sibirtsev8,Sibirtsev9,Sibirtsev10}.
Anyway, for our purpose the semiquantitative description of the
data on $\gamma{p}{\to}K^+\Lambda$(1520) as given by the solid line
is sufficient and, therefore, we refrain from exploring further 
options at this stage. 

Once the amplitude for $\gamma{p}{\to}K^+\Lambda$(1520) is established
we can then construct the reaction amplitude ${\mathcal M}$ for 
$\gamma{p}{\to}K^+K^-p$. It is given by~\cite{Collins}
\begin{eqnarray}
{\mathcal M}(s,t_1,\theta,\phi){=}\! 
\sum_{\lambda_2} \!M^{\lambda_1\lambda_a}_{\lambda_2\lambda_b}(s,t_1)G_2(s_2)
M^{\lambda_2}_{\lambda_c\lambda_d}(\theta,\phi),
\label{decay}
\end{eqnarray}
where $M^{\lambda_1\lambda_a}_{\lambda_2\lambda_b}$ is the 
production helicity amplitude given by Eq.~(\ref{hel}) and
$M^{\lambda_2}_{\lambda_c\lambda_d}$ is the amplitude corresponding to
the decay of the resonant state $2$ with helicity $\lambda_2$
into the system $cd$ with helicities $\lambda_c$ and $\lambda_d$,
respectively.  
Furthermore, 
$G_2$ is the propagator of the resonant state and $s_2$ is the 
invariant mass of the $2{\to}cd$ decay products.
The propagator is parametrized in a Breit-Wigner
form with an energy dependent width as proposed by Jackson~\cite{Jackson}, 
\begin{eqnarray}
G_2(s_2)=\frac{m_2\Gamma(\sqrt{s_2})}{s_2-m_2^2+im_2
\Gamma(\sqrt{s_2})} \ ,
\label{Wigner}
\end{eqnarray}
where $m_2$ is the resonance mass and the energy variation
of the width is given by 
\begin{eqnarray}
\Gamma(\sqrt{s_2})=\Gamma_0 
\left[\frac{\lambda^{1/2}(s_2,m_c^2,m_d^2)}
{\lambda^{1/2}(m_2^2,m_c^2,m_d^2)}\right]^{2l+1}\!\!\
\!\!\frac{\rho(\sqrt{s_2})}{\rho(m_2)} \ .
\end{eqnarray}
In the latter formula
$l$ is the orbital angular momentum of the $2{\to}cd$ decay
and $\Gamma_0$ is the width at $\sqrt{s_2}{=}m_2$, while 
$\rho(\sqrt{s_2})$ is a factor varying slowly with energy given by
Glashow and Rosenfeld as~\cite{Glashow}
\begin{eqnarray}
\rho(\sqrt{s_2})=\frac{1}{\sqrt{s_2}}\left[X^2{+}\frac{
\lambda(s_2,m_c^2,m_d^2)}{4s_2}\right]^{-l}
\end{eqnarray}
with the parameter $X{=}350$~MeV fixed from a fit of the baryonic 
resonances in the context of unitary symmetry.

\end{document}